\documentclass[reprint, pre, twocolumn, aps, showpacs, superscriptaddress]{revtex4-2}

\usepackage{graphicx}
\usepackage{dcolumn}
\usepackage{bm}
\usepackage[colorlinks= true, linkcolor=blue, citecolor=blue, urlcolor=blue]{hyperref}
\usepackage{lipsum}
\usepackage{mathtools}
\usepackage{xcolor}
\usepackage{multirow}
\usepackage{xcolor}
\usepackage{bbm}
\usepackage{xr}
\usepackage{mathtools}
\usepackage{amsthm, amssymb, mathrsfs}
\usepackage{empheq}
\usepackage{cases}
\hypersetup{
    colorlinks,
    linkcolor={blue},
    citecolor={blue},
    urlcolor={blue}
}

\begin{document}

\title{Voter Turnouts Govern Key Electoral Statistics}

\author{Ritam Pal}
\email{ritam.pal@students.iiserpune.ac.in}
\affiliation{Department of Physics, Indian Institute of Science Education and Research, Pune 411008, India.}
\author{Aanjaneya Kumar}
\email{aanjaneya@santafe.edu}
\affiliation{Department of Physics, Indian Institute of Science Education and Research, Pune 411008, India.}
\affiliation{Santa Fe Institute, 1399 Hyde Park Road, Santa Fe, NM 87501, USA}
\affiliation{High Meadows Environmental Institute, Princeton University, Princeton, NJ, 08544, USA}
\author{M. S. Santhanam}
\email{santh@iiserpune.ac.in}
\affiliation{Department of Physics, Indian Institute of Science Education and Research, Pune 411008, India.}

\begin{abstract}
Elections, the cornerstone of democratic societies, are usually regarded as unpredictable due to the complex interactions that shape them at different levels. In this work, we show that voter turnouts contain crucial information that can be leveraged to predict several key electoral statistics with remarkable accuracy. Using the recently proposed random voting model, we analytically derive the scaled distributions of votes secured by winners, runner-ups, and margins of victory, and demonstrating their strong correlation with turnout distributions. By analyzing Indian election data -- spanning multiple decades and electoral scales -- we validate these predictions empirically across all scales, from large parliamentary constituencies to polling booths. Further, we uncover a surprising scale-invariant behavior in the distributions of scaled margins of victory, a characteristic signature of Indian elections. Finally, we demonstrate a robust universality in the distribution of the scaled margin-to-turnout ratios.
\end{abstract}

\maketitle

The institution of election plays a pivotal role in every functioning democracy to ensure that the governing bodies are based on the people's mandate. While the rules for choosing candidates are often simple at the microscopic level, the effects arising from complex interactions among individuals can make electoral processes and their final outcomes unpredictable on larger scales. To address these complexities, tools from statistical physics and complex systems were applied to analyze and uncover patterns in electoral outcomes \cite{galam1999application, gelman2002mathematics, brams2008, CasForLor2009, galam2012, ForMacRed2013, SenCha2014, FerSucRam2014, BraDeA2017, Kon2019, redner2019reality, MigTor2020}.

Over the last few decades, aided by the availability of extensive election data, many earlier studies \cite{CosAlmAnd1999, ForCas2007, mantovani2011scaling, ChaMitFor2013, BokSzaVat2018, hosel2019universality} have attempted to identify universal patterns to characterize and simplify the complexities of electoral processes irrespective of microscopic details. While the distributions of vote shares garnered by candidates \cite{CalCroAnt2015, BurRanGir2016, MorHisNak2019, Kon2017} and voter turnouts\cite{BorBou2010, BorRayBou2012} have been extensively studied, they exhibit limited universality at best \cite{CosAlmAnd1999, ForCas2007, ChaMitFor2013}. Nevertheless, these distributions have proven valuable in flagging irregularities and detecting fraudulent practices in elections \cite{klimek2012statistical, jimenez2017testing, rozenas2017detecting}.

Among the many statistics of interest, the \emph{margin} of victory, defined as $M = V_w - V_r$, where $V_w$ and $V_r$ are the votes secured by the winner and runner-up, respectively, encodes key information about the competitiveness of elections. While margins of victory have been previously studied \cite{jacobson1987marginals, mccann1997threatening, mulligan2003empirical, magrino2011computing, xia2012computing, bhattacharyya2021predicting}, often independently of voter turnouts, our recent work \cite{pal2024universal} suggests that voter turnouts, in combination with margins, provide deeper insight into electoral dynamics. This extensive analysis of election data spread over many decades of elections held in $34$ countries demonstrated that the scaled distribution of the margin-to-turnout ratio exhibits a universal form \cite{pal2024universal}. Significant deviations from this universal behavior can indicate potential electoral malpractice \cite{pal2024universal, brigaldino2011elections, frear2014parliamentary, belarus_report, czwolek2021belarusian, bedford_2021}. Furthermore, we demonstrated that voter turnouts play a fundamental role in driving the margins and, together with a proposed random voting model (RVM), can accurately predict the scaled distribution of victory margins. This robust connection between turnouts and margins holds across multiple elections and electoral scales. These findings naturally raise the question: Can the distribution of other relevant electoral statistics be uncovered using voter turnout distributions?

To explore this question, large datasets covering a range of different electoral scales are required. India, the world's largest democracy, regularly conducts elections involving vast electorates (960 million in 2024), and its publicly available election data spans multiple decades and electoral scales. The diversity of India's linguistic and cultural landscape further adds to the complexities of its electoral outcomes, making it an excellent testing ground for assessing the robustness of the RVM framework.

In this Letter, using empirical data from Indian elections \cite{india_data, lokdhaba} spanning several decades and vastly different electoral scales, we demonstrate a strong correlation between the distributions of votes received by winners and runner-ups and voter turnouts. Leveraging this correlation and the RVM, we analytically predict the scaled distributions of the votes secured by winners and runner-ups using the corresponding turnout distributions. This prediction remarkably holds good at all the electoral scales -- from large parliamentary constituencies ($ \sim 10^6$ voters) down to the smallest polling booth levels ($\sim 10^2-10^3$ voters). Further, we show a rather surprising scale invariance of the margin distributions, a characteristic typical of Indian elections. Finally, a robust universality in the distribution of scaled margin-to-turnout ratio is demonstrated, strengthening our recent proposition.

We formalize our framework as follows: An {\it election} happens at all the $N$ electoral units following the first-post-the-past principle \cite{johnston2007politics}. Let the $i$-th electoral unit have $n^c_i$ candidates and $n^v_i$ eligible voters, where $i=1,2, \dots N$. Usually, only a fraction of the eligible voters cast their votes. This is termed the turnout $T_i \le n^v_i$. It is a direct indicator of the people's interest in the electoral process, and its distribution $g(T)$ encodes information about electoral statistics \cite{pal2024universal}. Let $v_{i,1}, v_{i,2}, \dots, v_{i,n^c_i}$ be the votes secured by $n^c_i$ candidates such that $\sum_j v_{i,j}=T_i$. The candidate securing the highest number of votes, $V_{w}$, is declared the winner, while the candidate with the second-highest votes, $V_{r}$, is the runner-up. By definition $V_w > V_r$. Further, the \emph{margin} of victory is defined as $M_i = V_{i, w} - V_{i, r}$, and indicates the extent of electoral competition.
\begin{table}[b]
\caption{\label{tab:example}Electoral units in various elections in India}
\begin{ruledtabular}
\begin{tabular}{llll}
Type & Type of electoral unit & Type of election & Size \\
\hline
GE-PB & Polling booth & General election  &  $10^3$ \\
GE-AC & Assembly constituency & General election  &  $10^5$ \\
GE-PC & Parliamentary constituency & General election  &  $10^6$ \\
SE-AC & Assembly constituency & State election  &  $10^5$ \\
\end{tabular}
\end{ruledtabular}
\label{table1}
\end{table}

\begin{figure}[t]
    \centering
    \includegraphics[width=\columnwidth]{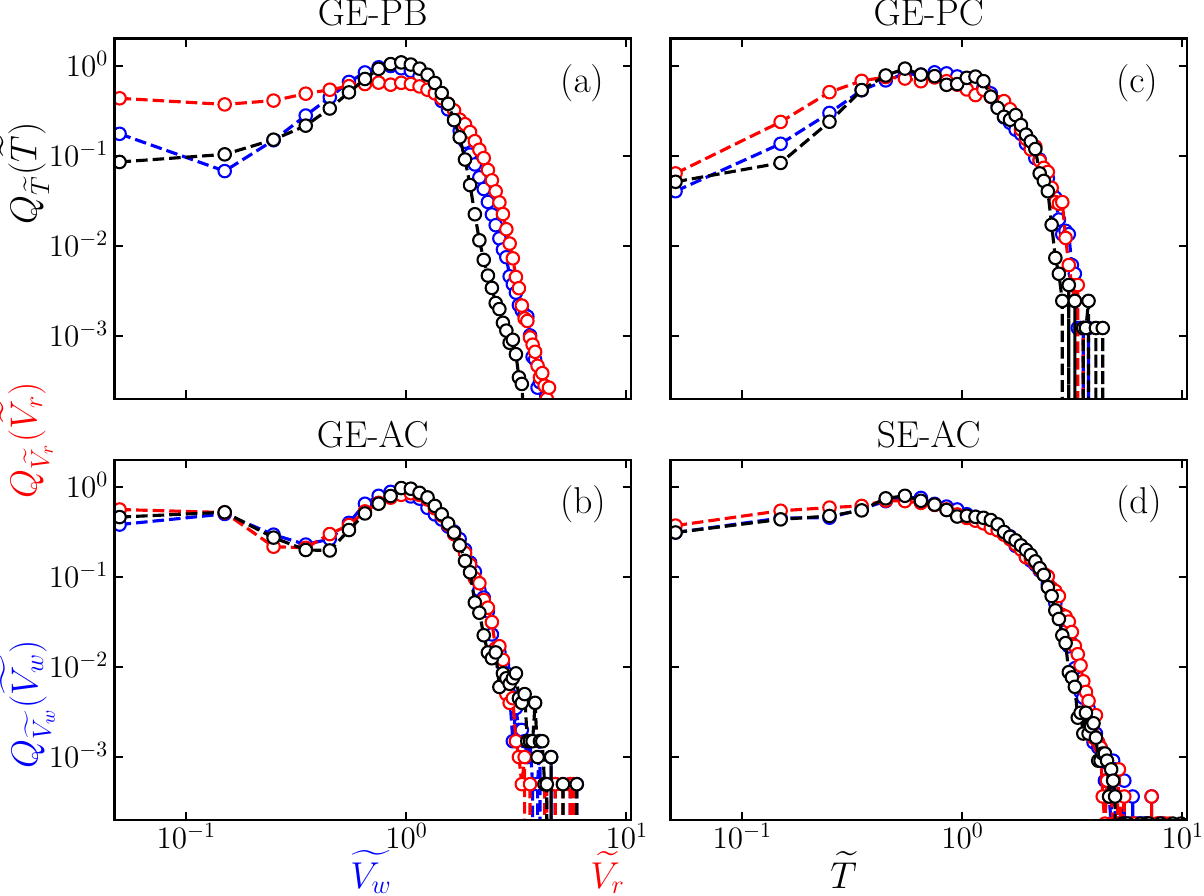}
    \caption{Winner, runner-up vote distributions, and turnout distributions, scaled by their respective means. Notably, at larger electoral scales (AC / PC), the winner and runner-up distributions mimic the corresponding turnout distribution.}
    \label{fig:1}
\end{figure}

The electoral units in this work have three distinct scales in terms of the size of the electorate: (i) parliamentary constituency (PC, largest scale), (ii) assembly constituency (AC, intermediate scale), and (iii) polling booth (PB, smallest scale). Table \ref{table1} describes the granularity of election data used in this work -- electoral units and their typical electorate size. While for the national-level general elections, we employ data from 3 different electoral scales (PC, AC, and PB), we use AC-level data for state elections. Depending on the electoral level considered, the winner and runner-up vote distributions have vastly different scales, with the winner vote distribution having wider support than the runner-up. However, when both distributions are scaled by their respective mean values (mean taken over all elections for which data is available), the winner and runner-up vote distributions -- $Q_{\widetilde{V_w}}(\widetilde{V_w})$ and $Q_{\widetilde{V_r}}(\widetilde{V_r})$ -- explicitly display a strong correlation with the corresponding scaled turnout distributions $Q_{\widetilde{T}}(\widetilde{T})$. Note that any variable $Y$, scaled by their mean $\langle Y \rangle$, is denoted by $\widetilde{Y} = Y / \langle Y \rangle$. Figure \ref{fig:1} displays this key result at four electoral units described in Table \ref{table1}. Remarkably, at larger electoral scales (PC and AC), not only the tail but the entire scaled distributions of winner and runner-up votes mimic the corresponding scaled turnout distribution as seen in Fig. \ref{fig:1}(b-d). This strong correlation indicates that the turnout distribution contains crucial information about different election statistics and can be leveraged to predict the scaled vote distributions of the winner and the runner-up. To explore this possibility, we employ our recently proposed random voting model (RVM) \cite{pal2024universal}, which is demonstrably effective at predicting the scaled distribution of several election statistics, such as the \emph{margin} of victory.

\begin{figure*}[t]
    \centering
    \includegraphics[width=\linewidth]{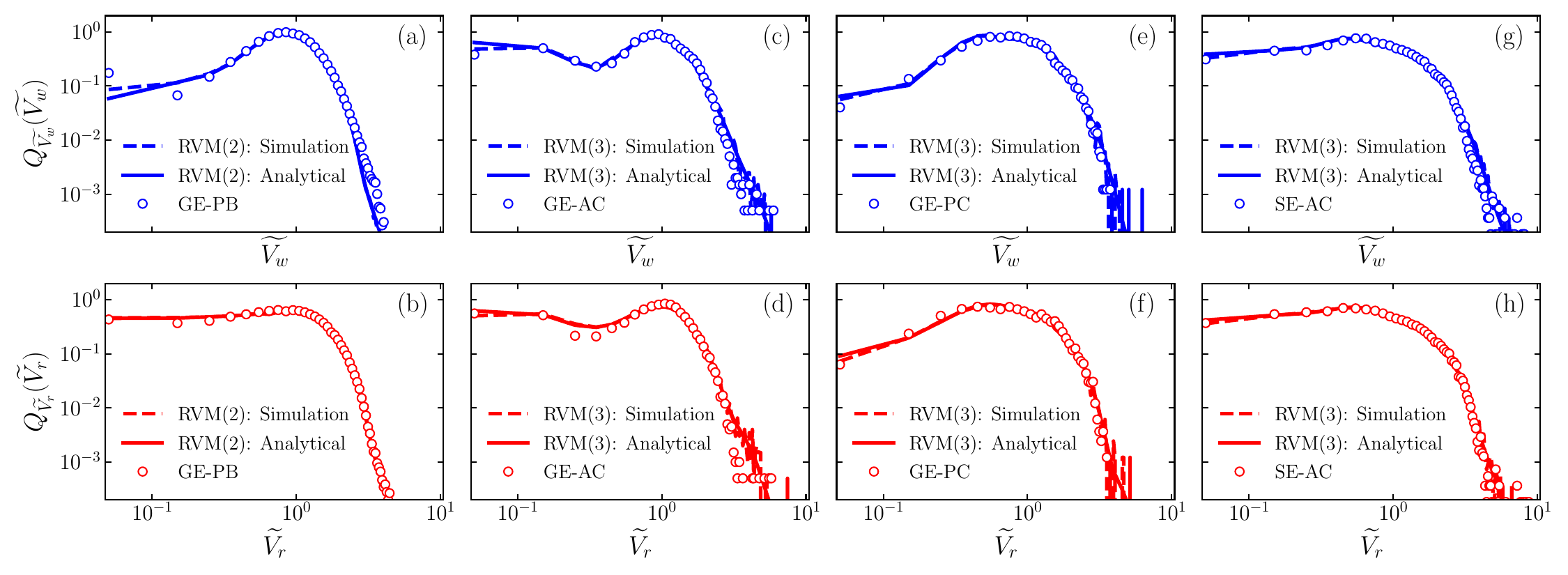}
    \caption{Winner and runner-up vote distributions scaled by their respective means. Panels (a, b), (c, d), and (e, f) depict, respectively, the scaled winner and runner-up vote distribution at the polling booth, assembly constituency, and parliamentary constituency level for Indian general elections. Panels (g, h) correspond to the distributions for the state elections at the assembly constituency level. The analytical predictions (solid lines) are in remarkable agreement with the empirical distributions (open circle). Predictions from RVM simulations (dashed line) closely follow the analytical curves.}
    \label{fig:2}
\end{figure*}

The random voting model, later denoted as {RVM $(T, n^c)$}, is built around the framework of elections described above and consists of $N$ electoral units with $n^c_i$ number of candidates contesting for the votes of $T_i$ electors who have voted in the $i$-th electoral unit. Then, the probability that $j$-th candidate attracts electors' votes is:
\begin{equation}
    p_{ij} = \frac{w_{ij}}{\sum_{k=1}^{n^c_i} w_{ik}}, ~~~w_{ij} \sim \mathcal{U}(0, 1), ~~ (j = 1, 2, \cdots n^c_i).
    \label{eq:prob}
\end{equation}
In this, $\mathcal{U}(0, 1)$ is the uniform distribution. This model was shown to capture the various statistical features of empirical election data, such as the margin distributions and the universal features embedded in the election data \cite{pal2024universal}. In particular, the random voting model with $n^c = 3$ candidates -- {RVM $(T, 3)$} -- predicts the scaled distribution of \emph{margin} remarkably well, irrespective of electoral scales and countries \cite{pal2024universal}. In this work, we employ a refined approach based on the notion of the effective number of parties defined as \cite{laakso1979effective}:
\begin{equation}
    ^{(E)}n^c_i = \frac{1}{\sum_{k=1}^{n^c_i} (V_{ik} / T_{i})^2}.
    \label{eq:effc}
\end{equation}
In large elections exercise such as that in India, even though many candidates join the fray, a few corner most of the votes. For instance, if all the votes are garnered by just one candidate, then $V_{i1}=T_i$, and $V_{ij}=0$ for $j=2, \dots n^c$. In this case, $^{(E)}n^c_i=1$. However, if all the votes are split equally among two candidates, $^{(E)}n^c_i=2$, thus, Eq. \ref{eq:effc} captures the idea of an effective number of candidates in $i$-th electoral unit. Further, by averaging over all the electoral units, we obtain: 
\begin{equation}
    ^{(E)}\Tilde{n}^c =  \left[\frac{1}{N}\sum_{k=1}^{N}{}^{(E)}n^{c}_k \right],
\end{equation}
where $\left[\:*\:\right]$ denotes the operation of extracting the closest integer value. And $^{(E)}\Tilde{n}^c $ indicates the effective number of candidates for an entire election at different electoral scales. From empirical data of Indian elections at different scales, we find $^{(E)}\Tilde{n}^c = 2$ at the polling booth (PB-GE) level for the General Elections. However, for all the other three cases of AC-GE, PC-GE, and AC-SE, we obtain $^{(E)}\Tilde{n}^c = 3$. Now, we shall solve RVM $(T, n^c)$ for $n^c = 2$ and $n = 3$ to analytically describe the results observed in Fig. \ref{fig:1}.\\

The primary object of interest is the distribution of the votes received by the winner and the runner-up. In the large turnout limit, $T >> 1$, the votes received by $j$-th candidate can be approximated as $V_j \approx p_j T$ (index $i$ is dropped as we focus on a single electoral unit). Consequently, the vote share is defined as:
\begin{equation}
    v_j = V_j/T ~~ \text{with} ~ j = 1, 2, \dots n^c.
    \label{eq:voteshare}
\end{equation}

Thus, in this limit, the vote share distribution is effectively the same as the distribution of $p_j$. Recall that $p_j$ can be constructed from random weights using Eq.~\ref{eq:prob}. These weights are $n^c$ \emph{i.i.d.} random variables $\{w_1, w_2, \cdots, w_{n^c}\}$ drawn from the uniform distribution $\mathcal{U}(0, 1)$. When arranged in ascending order, the random variable at the $k$-th place is defined as the $k$-th order statistics and is denoted by $w^{n^c}_{(k)}$ \cite{BarBalNag2008}. Specifically, $w^{n^c}_{(1)}$ and $w^{n^c}_{(n^c)}$ represent the smallest and the largest weights, respectively. Then, the joint probability distribution function (jpdf) or all the order statistics is given by:
\begin{equation}
    \mathbbm{P}\left(w^{n^c}_{(1)}, w^{n^c}_{(2)}, ... w^{n^c}_{(n^c)}\right) = n^c!.
    \label{eq:jpdf1}
\end{equation}
Finally, the winner's vote share $v_w$ and runner-up's vote share $v_r$ can be expressed in terms of order statistics as:
\begin{equation}
    v_w = \frac{w^{n^c}_{(n^c)}}{\sum_{k = 1}^{n^c}w^{n^c}_{(k)}} ~~~\text{and}~~~ v_r = \frac{w^{n^c}_{(n^c - 1)}}{\sum_{k = 1}^{n^c}w^{n^c}_{(k)}}.
    \label{eq:voteshare_order_stat}
\end{equation}

Their distributions can be obtained from the jpdf in Eq.~\ref{eq:jpdf1} by integrating out the other variables (for detailed calculations, see Supplementary Material \cite{supp}). When the number of candidates $n^c = 2$, the distribution of the winner's vote share $P_{v_w}(v_w)$ is found to be:
\begin{equation}
{P_{v_w}(v_w) = }
\begin{dcases}
     \frac{1}{v_w^2}, ~~\text{ if } ~~ \frac{1}{2} < v_w < 1,\\
     0, ~~\text{ otherwise}.
\end{dcases}
\label{eq:wdist1}
\end{equation}
The vote share distribution of the runner-up can also be calculated similarly and is given by:
\begin{equation}
{P_{v_r}(v_r) = }
\begin{dcases}
     \frac{1}{(1- v_r)^2}, ~~\text{ if } ~~ 0 < v_r \leq \frac{1}{2},\\
     0, ~~\text{ otherwise}.
\end{dcases}
\label{eq:rdist1}
\end{equation}
The conditions in Eqs.~\ref{eq:wdist1}-\ref{eq:rdist1} reflect the intuitive idea that when there are only two candidates, the winner's vote share cannot be less than $1/2$, and the runner-up cannot exceed $1/2$. For $n^c = 3$, the winner's vote share distribution is:
\begin{equation}
P_{v_w}(v_w) = 
\begin{dcases}
     \frac{3v_w - 1}{ v_w^3}, ~~\text{ if } ~~\frac{1}{3} < v_w \leq \frac{1}{2}\\
     \frac{1 - v_w}{v_w^3}, ~~\text{ if } ~~\frac{1}{2} < v_w < 1\\
     0, ~~\text{ otherwise},
\end{dcases}
\end{equation}
and the distribution of the runner-up's vote share is:
\begin{equation}
{P_{v_r}(v_r) = }
\begin{dcases}
    \frac{v_r (2 - 3 v_r)}{(1 - v_r)^2 (1 - 2 v_r)^2}, ~~\text{ if } ~~ 0 < v_r \leq \frac{1}{3}\\
    \frac{1 - 2 v_r}{v_r^2(1 - v_r)^2}, ~~\text{ if } ~~ \frac{1}{3} < v_r \leq \frac{1}{2}\\
     0, ~~ \text{ otherwise}.
\end{dcases}
\end{equation}
The vote share distribution of the winner and the runner-up are defined as piecewise functions. They are non-zero when $\frac{1}{3} < v_w < 1$ and $0 < v_r \leq \frac{1}{2}$, respectively.
Based on Eq.~\ref{eq:voteshare}, the distribution of unscaled variables $Y = (V_w, V_r)$, given $T$, is related to the scaled variables $y = (v_w, v_r)$ via:
\begin{equation}
    \mathcal{P}(Y|T) = \frac{1}{T}P_y\left(\frac{Y}{T}\right),
    \label{eq:PYT}
\end{equation}
where $P_y(y)$ is the probability density function for the scaled variable, $y = (v_w, v_r)$. The distribution of $Y$ for arbitrary turnout distribution $g(T)$ is:
\begin{equation}
    Q_Y(Y) = \int g(T)~\mathcal{P}(Y|T)~dT, 
    \label{eq:QY}
\end{equation}
with $\langle Y \rangle = \int Y ~  Q_Y(Y)~dY$.
Finally, for the scaled variable $\widetilde{Y} = Y / \langle Y \rangle$, the distribution is:
\begin{equation}
    Q_{\widetilde{Y}}(\widetilde{Y}) =  \langle Y \rangle ~ Q_Y(\widetilde{Y}  \langle Y \rangle). 
    \label{eq:QYscaled}
\end{equation}
\begin{figure}[h!]
    \centering
    \includegraphics[width=1\linewidth]{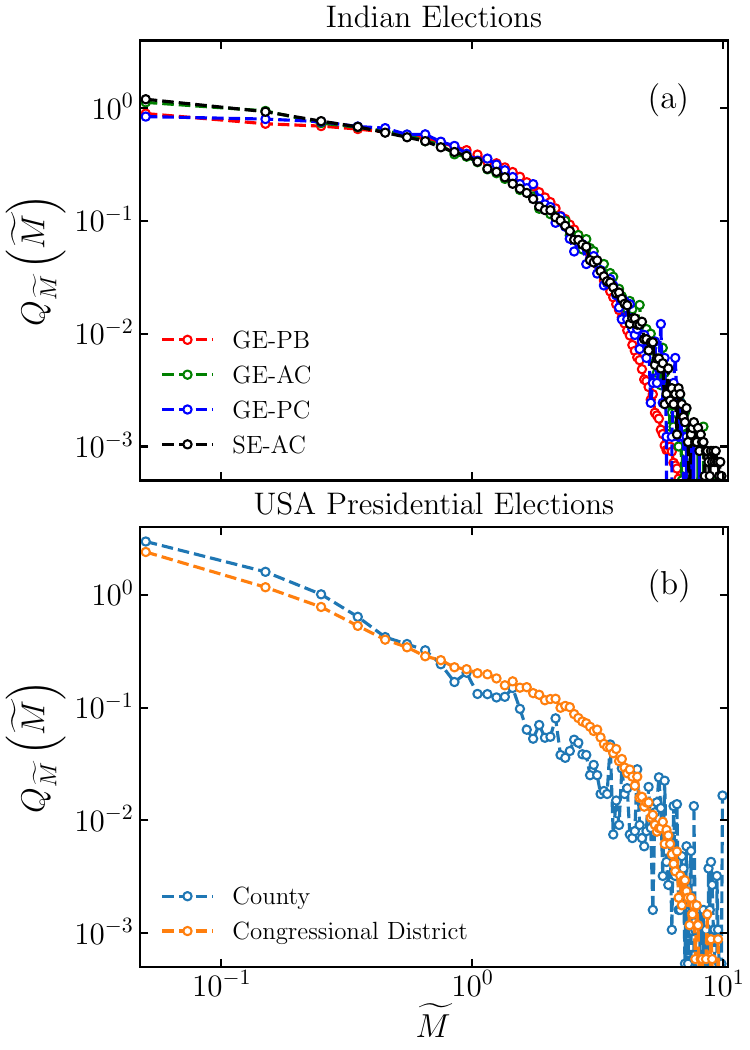}
    \caption{Margin distributions scaled by their respective means. (a) Data collapse in the scaled margin distributions of Indian elections at four electoral scales. (b) In contrast, such collapse is absent in the election data from the USA.}
    \label{fig:3}
\end{figure}
Using the empirical turnout distribution $g(T)$ from election data, Eq.~\ref{eq:QY} is numerically integrated. The resulting distribution is then scaled using Eq.~\ref{eq:QYscaled} to obtain the scaled distributions for the winner and runner-up vote shares, $ Q_{\widetilde{V_w}}(\widetilde{V_w})$ and $Q_{\widetilde{V_r}}(\widetilde{V_r})$, respectively. As demonstrated in Fig.~\ref{fig:2}, the analytical prediction (solid lines) is remarkably consistent with the empirical vote share distributions. The predictions from RVM simulations, which use the raw turnout data and $n^c = {^{(E)}\Tilde{n}^c}$ as inputs, closely follow the analytical distributions in Fig.~\ref{fig:2}. The scaled distributions of winner and runner-up votes depicted across all electoral scales, in Fig.~\ref{fig:2}, typically exhibit a power-law behavior in the tails for $\widetilde{V}_w, \widetilde{V}_r \gg 1$. Conversely, for $\widetilde{V}_w, \widetilde{V}_r \ll 1$, the distributions display different profiles. Remarkably, these differences are well captured by RVM predictions: RVM $(T, 2)$ accurately predicts distribution at the GE-PB level, while RVM $(T, 3)$ closely matches the distributions at the GE-AC, GE-PC, and SE-AC levels. Hence, the effective number of candidates (Eq.~\ref{eq:effc}) and the turnout distribution $g(T)$, when used within the RVM framework, successfully predict the winner and runner-up vote share distributions across distinct electoral scales.

Next, we consider the effect of voter turnouts $T$ on the \emph{margin} of victory $M$. To do this, firstly we define {\it specific margin} as $\mu = M / T = (V_w - V_r) / T$. In the large turnout $(T >> 1)$ limit, the specific margin can be expressed in terms of the order statistics of $w$ as:
\begin{equation}
    \mu = \frac{w^{n^c}_{(n^c)} - w^{n^c}_{(n^c - 1)}}{\sum_{k = 1}^{n^c}w^{n^c}_{(k)}},
\end{equation}
where $n^c$ is the number of candidates. Using the jpdf in Eq. \ref{eq:jpdf1}, for RVM $(T, 2)$ the distribution of specific margin can be obtained as (see Ref.~\cite{supp} for more details):
\begin{equation}
    P_{\mu}(\mu) = \frac{2}{(1 + \mu)^2},
\end{equation}
and for RVM $(T, 3)$, the distribution becomes:
\begin{equation}
    P_{\mu}(\mu) = \frac{(1 - \mu)(5 + 7\mu)}{(1 + \mu)^2(1 + 2\mu)^2}.
    \label{eq:24}
\end{equation}
Using Eq.~ \ref{eq:QY} - \ref{eq:QYscaled} and the empirical turnout distributions $g(T)$, the scaled margin distribution $Q_{\widetilde{M}}(\widetilde{M})$ can be obtained for the four different electoral scales which match the corresponding empirical distributions closely (see Ref.~\cite{supp}). Remarkably, the scaled distributions for the margin for Indian elections at four different scales collapse onto a single curve, as shown in Fig.~\ref{fig:3}(a). This data collapse is a direct consequence of the similarity in tail behavior in the corresponding turnout distributions (see Fig. \ref{fig:1}). This appears to be a characteristic of Indian elections and is not observed in most other countries. For instance, this data collapse is absent in the US elections for the empirical data at the County and Congressional district levels, as demonstrated in Fig.~\ref{fig:3} (b).
\begin{figure}[t]
    \centering
    \includegraphics[width=1\linewidth]{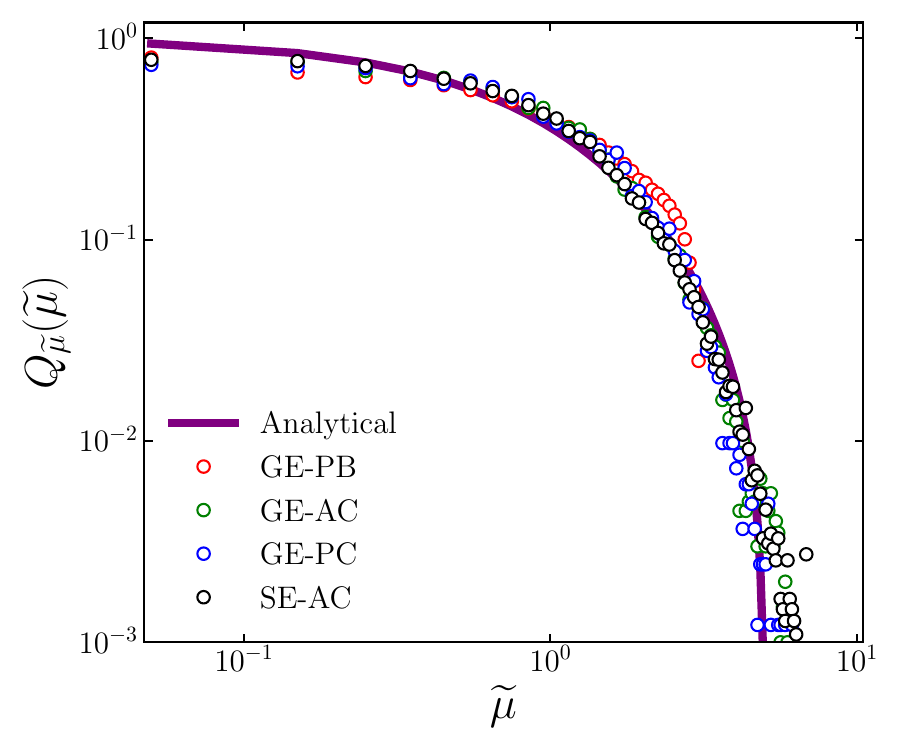}
    \caption{Specific margin distributions scaled by their respective means at distinct electoral scales in different types of Indian elections. The scaled specific margin distributions collapse on the analytical universal curve.}
    \label{fig:4}
\end{figure}

Recently, we have shown that the scaled distribution of specific margins $Q_{\widetilde{\mu}}(\widetilde{\mu})$, is universal irrespective of electoral scales and countries \cite{pal2024universal}. This universal distribution can be analytically obtained by rescaling Eq.~\ref{eq:24} with $\langle \mu \rangle = {1} / {2}+\ln \left({9 \sqrt[4]{3}} / {16}\right)$ (see Ref.~\cite{supp}). The empirical distributions from large election datasets are proposed to exhibit a better collapse to the analytical curve, as finite-size effects are suppressed. The presence of such universality helps us distill the complexity of the electoral processes in terms of universal behavior. As the electorate size in India is large (960 million in 2024) and datasets are available at various electoral scales, Indian elections provide the best test-bed to demonstrate such universality. Fig.~\ref{fig:4} demonstrates such universality. The distributions $Q_{\widetilde{\mu}}(\widetilde{\mu})$ at four distinct electoral scales (colored open circles) remarkably collapse onto the analytically predicted universal curve (solid line). This excellent collapse strengthens the proposition that in fairly conducted elections, the apparent deviation from universality can originate from finite data size.

In summary, elections are a source of excellent datasets for exploring collective decision-making by millions of people (interacting agents) at {\it distinct electoral scales}. However, most of the earlier works on elections have ignored the effects arising from differences in electoral scales. The voter turnout distribution $g(T)$ is a good indicator of the public trust and interest in the electoral process. Remarkably, we show that they also encode information about several crucial election statistics. Aided by election data from India -- the largest democracy in the world -- we demonstrate our results for different types of elections at multiple {\it distinct electoral scales}. Given the empirical $g(T)$ and an effective number of candidates for each electoral scale, we analytically obtain the scaled distributions of the winner and the runner-up votes using the framework of our recently proposed random voting model. The analytical predictions and the simulations are in excellent agreement with the empirical election data. Further, we demonstrate that the random voting model is effective in predicting the scaled margin distributions of Indian elections at vastly different electoral scales. Surprisingly, the scaled margin distribution remains invariant with respect to changes in electoral scales, making it a characteristic feature of Indian elections. Finally, the scaled specific margin distributions of Indian elections show a remarkable data collapse, strengthening the recently proposed universality \cite{pal2024universal}. This work paves the way for a wider understanding of electoral statistics from turnout distributions.

\begin{acknowledgments}
R.P. and A.K. thank the Prime Minister's Research Fellowship of the Government of India for financial support. M.S.S. acknowledges the support of a MATRICS Grant from SERB, Government of India, during the early stages of this work. The authors acknowledge the National Supercomputing Mission for the use of PARAM Brahma at IISER Pune.
\end{acknowledgments}

\newpage
\setcounter{page}{1}
\renewcommand{\thepage}{S\arabic{page}}
\setcounter{equation}{0}
\renewcommand{\theequation}{S\arabic{equation}}
\setcounter{figure}{0}
\renewcommand{\thefigure}{S\arabic{figure}}
\setcounter{section}{0}
\renewcommand{\thesection}{S\arabic{section}}
\setcounter{table}{0}
\renewcommand{\thetable}{S\arabic{table}}

\onecolumngrid
\newpage
\begin{center}
\textbf{\large Supplemental Material for ``Voter Turnouts Govern Key Electoral Statistics"}
\end{center}

This Supplemental Material provides further discussion and derivations which support the findings reported in the Letter, and provides details of the models and simulations used to validate the results. 

\tableofcontents

\section{Description of Random Voting Model (RVM)}
\label{sec:1}
\noindent In the Random Voting Model, $N$ electoral units are considered, with $n^c_i$ number of candidates contesting to win votes from $T_i$ voters present to cast their votes in the $i$-th electoral unit. Each of the $n^c_i$ candidates is assigned a random weight $w_{ij}$. These weights are drawn independently from a uniform distribution between $0$ and $1$. The corresponding probability $p_{ij}$ of receiving votes is calculated by normalizing these weights. Hence, we have the following,
\begin{equation}
    w_{ij} \sim \mathcal{U}(0, 1) \text{ and } p_{ij} = \frac{w_{ij}}{\sum_{k=1}^3 w_{ik}}; \text{ with } j = 1, 2\cdots n^c_i.
    \label{eq:prob-def}
\end{equation}
For the rest of the analysis, we focus on a single ($i$-th) electoral unit with voter turnout $T$ and drop the corresponding index $i$ for brevity. Hence,
\mathtoolsset{centercolon}
\begin{equation}
    w_{ij} := w_j \text{ and } p_{ij} := p_j.
\end{equation}

\section{Analytical Calculations of Different Election Statistics}
\noindent For large turnout $(T \gg 1)$, it is reasonable to assume the number of votes received by $j$-th candidate is proportional to their probability $p_j$, in particular, $V_j \approx p_j T$. Hence, for $T \gg 1$, the votes received by the winner  $V_w$ can be approximated as,
\begin{equation}
V_w \approx p_{max}T,
\end{equation}
and the votes received by the runner-up $V_r$ as,
\begin{equation}
V_r \approx p_{2nd\:max}T,
\end{equation}
and the margin $M = V_w - V_r$ can also be written as,
\begin{equation}
M \approx \left( p_{max} - p_{2nd\:max}\right)T,
\end{equation}
where $p_{max}$ and $p_{2nd \:max}$ correspond to the largest and the second largest probabilities assigned to the candidates. For example, if the number of candidates $n^c = 3$ and the probabilities $p_1, p_2,$ and $p_3$ assigned to those 3 candidates are $0.2, 0.5,$ and $0.3$, then $p_{max} = p_2 = 0.5$ and $p_{2nd \:max} = p_3 = 0.3$. 

\subsection{Order statistics and its connection to winner and runner-up vote share and margins}
\noindent Consider $n$ \emph{iid} random variables $\{X_1, X_2 \dots X_n\}$ drawn from a distribution $\rho(x)$. When arranged in ascending order, the random variable at the $k$-th spot is defined as the $k$-th order statistics. In particular, $n$-th and $1$-st order statistics correspond to the maximum and minimum of those $n$ random variables, respectively. The $k$-th order statistics of the random variable $X$ is denoted by $X_{(k)}$.\\

\noindent The joint probability density of all the order statistics of the above-mentioned $n$ random variables, $\mathbbm{P}\left(x_{(1)}, x_{(2)}, ... x_{(n)}\right)$, defined as the probability density that the random variable $X_{(k)}$ takes the value $x_{(k)}$ for $k \in \{ 1, 2, \dots, n\}$, is
\begin{equation}
    \mathbbm{P}\left(x_{(1)}, x_{(2)}, ... x_{(n)}\right) = n!\prod_{k=1}^{n}\rho\left(x_{(k)}\right).
\end{equation}

\noindent Now as described in Eq.~\ref{eq:prob-def} the probabilities $p_j$ can be expressed in terms of $w_j$. Hence the winner vote share $v_w = V_w / T$, runner-up vote share $v_r = V_r / T$ and the specific margin $\mu = M / T$ can be expressed in terms of $w$ as the following,
\begin{center}
\begin{align}
    v_w &= \frac{V_w}{T} \approx p_{max} = \frac{w_{max}}{\sum_{k = 1}^{n^c}w_{k}} = \frac{w_{(n^c)}}{\sum_{k = 1}^{n^c}w_{(k)}}\\
    \label{eq:v_w}
    v_r &= \frac{V_r}{T} \approx p_{2nd\: max} = \frac{w_{2nd\:max}}{\sum_{k = 1}^{n^c}w_{k}} = \frac{w_{(n^c - 1)}}{\sum_{k = 1}^{n^c}w_{(k)}}\\
    \label{eq:v_r}
    \mu &= \frac{M}{T} \approx p_{max} - p_{2nd\: max} = \frac{w_{max} - w_{2nd\:max}}{\sum_{k = 1}^{n^c}w_{k}} = \frac{w_{(n^c)} - w_{(n^c - 1)}}{\sum_{k = 1}^{n^c}w_{(k)}},
\end{align}
\end{center}
where $w_{(k)}$ is the $k$-th order statistics \cite{BarBalNag2008}.

\subsection{Random Voting Model with two candidates}
\noindent In the two-candidate Random Voting Model, we have $n = n^c = 2$ and $p(x) = \mathcal{U}(0, 1)$. Hence, the joint probability distribution of all the order statistics have the following form,
\begin{center}
    \begin{align}
        \mathbbm{P}\left(w_{(1)}, w_{(2)}\right) = 2! = 2; \text{ with } 0<w_{(1)}<w_{(2)}<1,
        \label{eq:jpd-2}
    \end{align}
\end{center}
and $\mathbbm{P}\left(w_{(1)}, w_{(2)}\right) = 0$ otherwise, with the following normalization:
\begin{equation}
    \int_{0}^{1}dw_{(2)}\int_{0}^{w_{(2)}}2 dw_{(1)} = 1.
\end{equation}

\subsubsection{Winner vote share distribution}
\noindent From the joint probability distribution of all the order statistics (Eq.~\ref{eq:jpd-2}), the approximate vote share distribution of the winner can be obtained as,
\begin{center}
    \begin{align}
        \nonumber P_{v_w}\left(v_w\right) & = 2 \nonumber \int_{0}^{1}dw_{(2)}\int_{0}^{w_{(2)}}\delta\left(v_w - \frac{w_{(2)}}{w_{(1)} + w_{(2)}}\right)dw_{(1)},\\
        & = 2 \int_{0}^{1}\frac{2 w_{(2)}}{v_w^2} \nonumber \mathbbm{1}_{1 / 2 \leq v_w < 1} dw_{(2)},\\
    \end{align}
\end{center}
or,
\begin{numcases}{P_{v_w}(v_w) = }
     \frac{1}{v_w^2} \text{ if } \frac{1}{2} \leq v_w < 1\\
     0, \text{ otherwise}.
\end{numcases}

\subsubsection{Runner-up vote share distribution}
\noindent We can similarly calculate the probability density function of the runner-up vote share as the following,
\begin{center}
    \begin{align}
        \nonumber P_{v_r}\left(v_r\right) & = 2 \nonumber \int_{0}^{1}dw_{(2)}\int_{0}^{w_{(2)}}\delta\left(v_r - \frac{w_{(1)}}{w_{(1)} + w_{(2)}}\right)dw_{(1)},\\
        & = 2 \int_{0}^{1}\frac{2 w_{(2)}}{(1 - v_r)^2} \nonumber \mathbbm{1}_{0 < v_r < 1 / 2} dw_{(2)},\\
    \end{align}
\end{center}
or,
\begin{numcases}{P_{v_r}(v_r) = }
     \frac{1}{(1 - v_r)^2} \text{ if } 0 < v_r < \frac{1}{2}\\
     0, \text{ otherwise}.
\end{numcases}

\subsubsection{Specific margin distribution}
\noindent Similarly, the distribution of the specific margin $\mu = M / T$ can be obtained as,
\begin{center}
    \begin{align}
        \nonumber P_{\mu}\left(\mu\right) & = 2 \nonumber \int_{0}^{1}dw_{(2)}\int_{0}^{w_{(2)}}\delta\left(\mu - \frac{w_{(2)} - w_{(1)}}{w_{(1)} + w_{(2)}}\right)dw_{(1)},\\
        & = 2 \int_{0}^{1}\frac{4 w_{(2)}}{(1 + \mu)^2} \nonumber dw_{(2)},\\
    \end{align}
\end{center}
or,
\begin{numcases}{P_{\mu}(\mu) = }
     \frac{2}{(1 + \mu)^2} \text{ if } 0 < \mu < 1\\
     0, \text{ otherwise}.
\end{numcases}

\subsection{Random Voting Model with three candidates}
\noindent In the three-candidate Random Voting Model, we have $n = n^c = 3$ and $p(x) = \mathcal{U}(0, 1)$. Then, the joint probability distribution of all the order statistics is,
\begin{center}
    \begin{align}
        \mathbbm{P}\left(w_{(1)}, w_{(2)}, w_{(3)}\right) = 3! = 6; \text{ with } 0<w_{(1)}<w_{(2)}<w_{(3)}<1,
    \end{align}
\end{center}
and $\mathbbm{P}\left(w_{(1)}, w_{(2)}, w_{(3)}\right) = 0$ otherwise, with the following normalization:
\begin{equation}
    \int_{0}^{1}dw_{(3)}\int_{0}^{w_{(3)}}dw_{(2)}\int_{0}^{w_{(2)}} 6 dw_{(1)} = 1.
\end{equation}
\subsubsection{Winner vote share distribution}
\noindent From the joint probability distribution of all the order statistics, we calculate the approximate probability density function of the winner vote share $v_w = V_w / T$ as follows, 
\begin{center}
    \begin{align}
        \nonumber P_{v_w}\left(v_w\right) & = 6 \nonumber \int_{0}^{1}dw_{(3)}\int_{0}^{w_{(3)}}dw_{(2)}\int_{0}^{w_{(2)}} \delta\left(v_w - \frac{w_{(3)}}{w_{(1)} + w_{(2)} + w_{(3)}}\right)dw_{(1)},\\
        & = 6 \int_{0}^{1}dw_{(3)}\int_{0}^{w_{(3)}} \frac{w_{(3)}}{v_w^2} \nonumber \mathbbm{1}_{0<\frac{w_{(3)} - v_w \left(w_{(2)} + w_{(3)}\right)}{v_w}<w_{(2)}} dw_{(2)},\\
    \end{align}
\end{center}
or,
\begin{numcases}{P_{v_w} = }
     6 \int_{0}^{1} w_{(3)}^2\frac{3v_w - 1}{2 v_w^3}dw_{(3)}, \text{ if } \frac{1}{3} < v_w \leq \frac{1}{2}\\
     6 \int_{0}^{1} w_{(3)}^2\frac{1 - v_w}{2 v_w^3}dw_{(3)}, \text{ if } \frac{1}{2} < v_w \leq 1\\
     0, \text{ otherwise}.
\end{numcases}

\noindent Finally, after performing the integral, we get
\begin{numcases}{P_{v_w} = }
     \frac{3v_w - 1}{ v_w^3} \text{ if } \frac{1}{3} < v_w \leq \frac{1}{2}\\
     \frac{1 - v_w}{v_w^3}, \text{ if } \frac{1}{2} < v_w < 1\\
     0, \text{ otherwise}.
\end{numcases}

\subsubsection{Runner-up vote share distribution}
\noindent Similarly, the probability density function of the runner-up vote share $v_r = V_r / T$ can be obtained as follows,

\begin{center}
    \begin{align}
        \nonumber P_{v_r}\left(v_w\right) & = 6 \nonumber \int_{0}^{1}dw_{(3)}\int_{0}^{w_{(3)}}dw_{(2)}\int_{0}^{w_{(2)}} \delta\left(v_r - \frac{w_{(2)}}{w_{(1)} + w_{(2)} + w_{(3)}}\right)dw_{(1)},\\
        & = 6 \int_{0}^{1}dw_{(3)}\int_{0}^{w_{(3)}} \frac{w_{(2)}}{v_r^2} \nonumber \mathbbm{1}_{0< (1 / v_r - 1) w_{(2)} - w_{(3)}< w_{(2)}} dw_{(2)},\\
    \end{align}
\end{center}
or,
\begin{numcases}{P_{v_r}(v_r) = }
     6 \int_{0}^{1} w_{(3)}^2\frac{v_r(2-3v_r)}{2 (1 - v_r)^2 (1 - 2v_r)^2}dw_{(3)}, \text{ if } 0 < v_r \leq \frac{1}{3}\\
     6 \int_{0}^{1} w_{(3)}^2\frac{1 - 2v_r}{2v_r^2 (1 - v_r)^2 }dw_{(3)}, \text{ if } \frac{1}{3} < v_r < \frac{1}{2}\\
     0, \text{ otherwise}.
\end{numcases}

\noindent Finally, after performing the integral, we get
\begin{numcases}{P_{v_r}(v_r) = }
    \frac{v_r (2 - 3v_r)}{(1 - v_r)^2 (1 - 2v_r)^2} \text{ if } 0 < v_r \leq \frac{1}{3}\\
     \frac{1 - 2v_r}{v_r^2(1 - v_r)^2}, \text{ if } \frac{1}{3} < v_r < \frac{1}{3}\\
     0, \text{ otherwise}.
\end{numcases}
\subsubsection{Specific margin distribution}
\noindent We obtain the distribution of specific margin $\mu = M / T$ as follows,
\begin{center}
    \begin{align}
        \nonumber P_{\mu}\left(\mu\right) & = 6 \nonumber \int_{0}^{1}dw_{(3)}\int_{0}^{w_{(3)}}dw_{(2)}\int_{0}^{w_{(2)}} \delta\left(\mu - \frac{w_{(3)} - w_{(2)}}{w_{(1)} + w_{(2)} + w_{(3)}}\right)dw_{(1)},\\
        & = 6 \int_{0}^{1}dw_{(3)}\int_{0}^{w_{(3)}} \frac{w_{(3)} - w_{(2)}}{\mu^2} \nonumber \mathbbm{1}_{0< \frac{w_{(3)} - w_{(2)} - \mu \left(w_{(3)} + w_{(2)}\right)}{\mu} < w_{(2)}} dw_{(2)},\\
    \end{align}
\end{center}
or,
\begin{numcases}{P_{\mu}(\mu) = }
     6 \int_{0}^{1}w_{(3)}^2 \frac{5 + 2\mu - 7\mu^2}{2 (1 + \mu)^2 (1 + 2\mu)^2}dw_{(3)}, \text{ if } 0 < \mu < 1\\
     0, \text{ otherwise}.
\end{numcases}

\noindent Finally, after performing the integral, we get
\begin{numcases}{P_{\mu}(\mu) = }
    \frac{(1 - \mu) (5 + 7\mu)}{(1 + \mu)^2 (1 + 2 \mu)^ 2} \text{ if } 0 < \mu < 1\\
    0, \text{ otherwise}.
\end{numcases}

\subsection{Calculating the \emph{scaled} distributions}
\noindent The winner and runner-up vote shares and specific margins are random variables scaled by the voter turnout $T$. However, through a simple change of variable, $Y = yT$, we can obtain the conditional distributions of the unscaled variables as,
\begin{equation}
    \mathcal{P}\left(Y|T\right) = \frac{1}{T}P_y\left(Y / T \right),
\end{equation}
where $y$ can be $v_w, v_r$, and $\mu$ and $Y$ represents unscaled variables  $V_w, V_r$, and $M$ respectively. The distribution of $Y$ for an arbitrary turnout distribution $g(T)$ can be obtained as,
\begin{equation}
    Q_Y(Y) = \int g(T) ~ \mathcal{P}(Y|T)dT, 
\end{equation}
with $\langle Y\rangle$ defined as,
\begin{equation}
    \langle Y \rangle = \int Y ~ Q_Y(Y)dY.
\end{equation}
\noindent Finally the distribution of \emph{scaled} $Y$, defined as $\widetilde{Y} = Y / \langle Y \rangle$, can be obtained as follows,
\begin{equation}
    {Q}_{\widetilde{Y}}(\widetilde{Y}) =  \langle Y \rangle ~ Q_{Y}(\widetilde{Y}  \langle Y \rangle)
\end{equation}
\noindent Again, the dummy random variable $Y$ can be either, $V_w, V_r$, and $M$.
\subsection{The universal distribution of scaled specific margins}
\noindent The distribution of scaled specific margins $(\widetilde{\mu} = \mu / \langle \mu \rangle)$ is universal and can be obtained through a simple scaling of the distribution of specific margins from 3-candidate RVM by its mean. The distribution ${Q}_{\widetilde{\mu}}(\widetilde{\mu})$ has the following form:
\begin{equation}
    {Q}_{\widetilde{\mu}}(\widetilde{\mu}) =  \langle \mu \rangle ~ Q_{\mu}(\widetilde{\mu}  \langle \mu \rangle) =  \frac{\langle \mu \rangle(1 - \widetilde{\mu} \langle \mu \rangle)(5 + 7~\widetilde{\mu} \langle \mu \rangle)}{(1 + \widetilde{\mu} \langle \mu \rangle)^2(1 + 2~\widetilde{\mu} \langle \mu \rangle)^2},
\end{equation}
with $\langle \mu\rangle = \frac{1}{2}+\ln \left(\frac{9 \sqrt[4]{3}}{16}\right)$.
\section{Simulation Details}
\noindent The Random Voting Model, $\mathcal{V}{\left(T, n^c\right)}$, with only voter turnouts and number of candidates as an input, can predict the distributions of the winner, runner-up votes, and the margins, when scaled appropriately. For the purpose of this model, the length of the voter turnout array can be assumed to be the total number of electoral units. The model can be simulated by drawing $n^c$ random numbers from a uniform distribution $\mathcal{U}(0, 1)$ for each electoral unit. These random numbers are further normalized using Eq.~\ref{eq:prob-def} to find the probabilities for attracting votes for each candidate. Next, each of the $T_i$ (voter turnout in $i$th electoral unit) electors cast their votes according to the previously calculated probabilities. Finally, all the votes in that unit are counted, and votes received by the winner and the runner-up, along with the margin of victory, are stored. This is repeated for all the electoral units to obtain arrays of the winner votes, runner-up votes, and margins.

\section{Prediction of Scaled Margin Distribution at Different Scales}
\begin{figure}[ht!]
    \centering
    \includegraphics[width=1\linewidth]{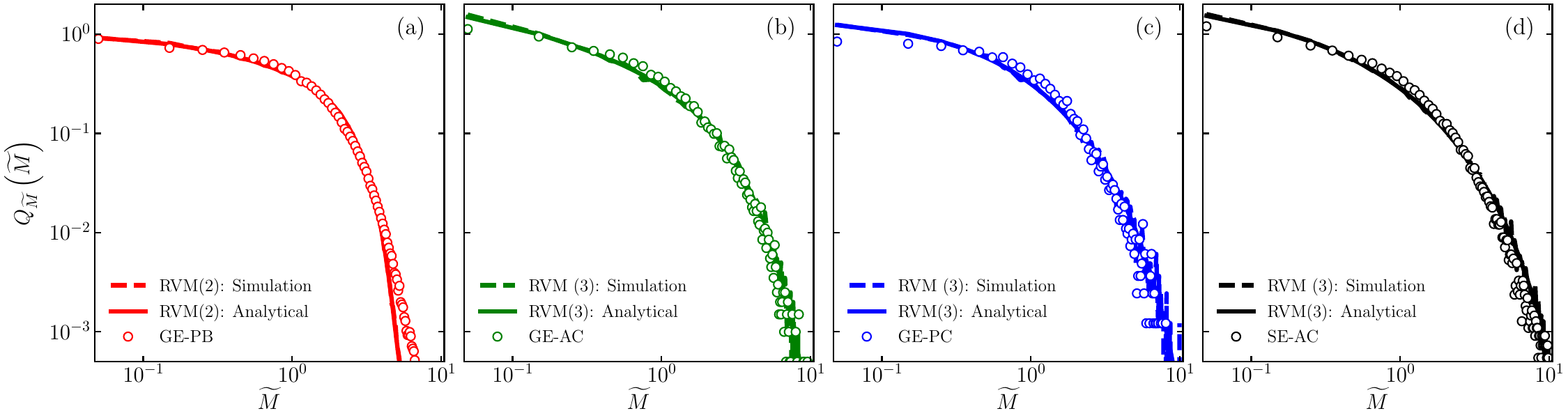}
    \caption{Scaled distribution of margins at different electoral scales. Panel (a) - (c) demonstrates the excellent prediction of scaled margin distribution for Indian General Elections at the polling booth, assembly constituency, and parliamentary constituency levels, respectively. Panel (d) demonstrates the same for state Assembly Elections at the assembly constituency level. The open circles denote the empirical distributions. The dashed and solid lines denote the prediction from RVM simulations and analytical calculations, respectively.}
    \label{fig:s1}
\end{figure}
\newpage
\section{Data Summary}
\noindent The parliamentary and assembly constituency level data of Indian General elections, along with the assembly constituency level data of the Assembly election, were obtained from the election data repository of Lok Dhaba \cite{lokdhaba}. The polling booth level data for the general elections were collected from the websites of chief electoral officers of different states in India \cite{india_data}. The following table contains the summary of Indian election data.
\begin{table}[h]
\centering
\begin{tabular}{|l|l|l|l|l|}
\hline
Election Type & General Election & General Election & General Election & State Election \\
\hline  
Electoral Scale & Parliamentary Constituency & Assembly Constituency & Polling Booth & Assembly Constituency \\
\hline
Time Span &  1962-2019& 1999-2019 & 2004-2019 & 1961-2023 \\
\hline
Number of Elections &  52 (including bye-elections)& 5 & 4 & 61\\
\hline
Average Turnout & 587329& 116577& 583& 86484\\
\hline
Average Winner Votes & 286807& 56874& 348& 39884\\
\hline
Average Runner-up Votes & 201281& 38887& 159& 28562\\
\hline
Average Margin & 85526& 17987& 189& 11322\\
\hline
\end{tabular}
\caption{The table presents typical values of voter turnout and winner votes, runner-up votes, and winning margins across different electoral levels for various types of Indian elections. The available data for the mentioned time spans were consolidated for each country and used to calculate the respective averages. The data for an electoral unit is considered valid if it meets the following criteria:  (a) a complete list of votes received by all candidates was available, (b) at least two candidates contested the election, and (c) the turnout was non-zero.}
\label{table}
\end{table}
\end{document}